# MOSFET GIDL Current Variation with Impurity Doping Concentration – A Novel Theoretical Approach

A. Sen and J. Das

*Abstract*—This paper depicts the actual variation of gate-induced-drain-leakage current with impurity doping concentration by complete qualitative and quantitative approach. De Casteljau's algorithm is applied to describe the band-to-band tunneling in a thin gate oxide n-MOSFET and the results are remarkably matched. Moreover for the very first time, the dependency of the leakage current over impurity density in the MOSFET drain region is explained in the context of pure geometrical approach. Surprisingly one of the proportionality constant exactly behaves like impurity gradient which results same characteristics as MOSFET Drain impurity doping profile measured using 2D simulator.

## I. Introduction

Drain Induced Gate Leakage is observed when gate overlaps drain in deep sub-micron. This phenomenon is the one of the most significant leakage mechanism which acts like a dead end in the journey of VLSI MOSFET scaling as well as lead to device breakdown sometimes. This adverse effect was frequently noticed in DRAM trench transistor cells [1]. Moreover, this mysterious leakage current was also popped up in the case of discharging of storage node and EEPROM memory cells with FETMOS structure. J. Chen and his team observed this leakage phenomenon while experimenting the $I_D$-$V_{DG}$ sub threshold characteristics for a 88 Å gate-oxide in n-mos [2]. Kuo-Feng, Ching-Yuan Wu, Ja-Hao Chen, Shyh-Chyi Wong and Yeong-Her Wang contributed various innovative theoretical models to predict the behavior of this leakage current and the results were well matched with the experiment. At last the hard work of Xiaoshi Jin, Xi Liu and Jong-Ho Lee designed an almost error free model [5] by calculating the net magnitude of involved electric field. Another well acceptable electric field expression for GIDL current using work function engineering calculated by Farkhanda Ana should be highlighted here. The prime objective of the previous researchers was to reveal the dependency of this leakage current over external voltage $V_{DG}$ and most of the theoretical and experimental results confirmed that $I_{GIDL}$ is varied from $10^{-19}$ to $10^{-8}$ Amp with a negative gate voltage variation from -8V to -2V. However, these results explain the approximate behavior of $I_{GIDL}$ but the way to minimize this unwanted leakage was not efficiently solved. Well, another strong parameter which controls $I_{GIDL}$ is impurity doping profile. According to Endoh's model $I_{GIDL}$ is suppressed in both very low and very high impurity doping concentration and amplified at moderate doping concentration ($\sim 10^{18}$ cm$^{-3}$). Interestingly both of the breakdown mechanism – Avalanche and Zener are dominating in very low and very high doping concentration respectively. Thus a trade-off is occurred between suppression of leakage mechanism and breakdown phenomenon. Due to this fact a close study regarding the dependency of $I_{GIDL}$ over impurity doping is essential and this paper serves us exactly the same.

## II. GIDL Mechanism

When MOSFET is scaled down significantly in sub-micron range Gate overlaps the drain region and an Oxide-Semiconductor contact is arisen in the context of reverse bias. With further increase of external voltage the existed depletion region surrounded over the drain is penetrated into the drain and a band bending is introduced. If this band bending exceeds the Si band gap the electrons participating in covalent bond is torn out and electron-hole pair is generated. This generated electron-hole pair introduces band-to-band tunneling and results GIDL current. For higher impurity doping concentration the penetration of depletion region into drain is prevented and after a critical value the drain becomes impenetrable and $I_{GIDL}$ is suppressed. On the other hand if impurity doping concentration is too low the depletion widths and tunneling barrier is too wide to produce $I_{GIDL}$. To make this leakage mechanism concept compatible to our proposed model we have to think in a slightly different approach. There are two main function which controls $I_{GIDL}$ generation: first one is band to band tunneling width and the second one is rate of electron-hole generation. It is evident that Impurity doping concentration is inversely and directly proportional to the first and second factor respectively. Figure-1 explains the details of GIDL phenomenon and dependency of impurity doping concentration.

This manuscript is submitted on.
A. Sen and J. Das are with Dept. of Physics, Jadavpur University, Kolkata – 700032, India.
Email: senarnesh.elec@gmail.com and jayoti.das@gmail.com

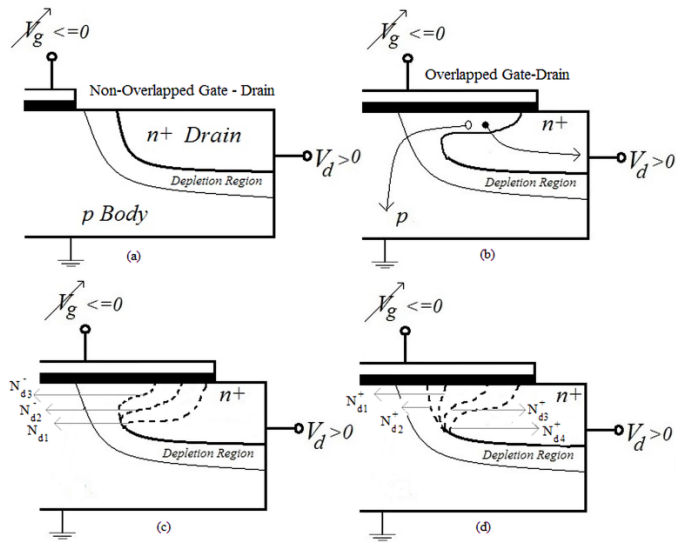

Fig-1: GIDL Mechanism (a) do not Occur in Non-Overlapped Gate-Drain (b) In Overlapped Gate-Drain. Region of Penetration variation with $N_d$ (c) increases ($N_{d1}^- < N_{d2}^- < N_{d3}^-$) and (d) decreases ($N_{d1}^+ > N_{d2}^+ > N_{d3}^+ > N_{d4}^+$).

### III. MODEL DEVELOPMENT

The axes are set as shown in Figure-2(b) where P1, P2 and P3 are the corresponding Bezier curve control points. These P1, P2 and P3 are the corresponding depletion width, Impurity doping concentration and bending potential related control points respectively. Depending upon their co-ordinates the equation of potential barrier is extracted as shown in Equation-(2) and by using WKB Approximation method the Band-to-Band Tunneling Probability (T) Expression is calculated. All the x axis parameters are normalized with respect to $W_d$ and denotes with suffix n. The normalized value of $x_2$ is $x_2/W_d = x_{n2}$. The coordinates of the classical tunneling points are related with the Bezier curve control point by a new factor $\lambda$ which is ratio as

$$x_{bn} = \frac{x_{n2}}{\lambda} = \frac{x_2}{W_d \lambda} \quad or, \quad x_2 = \lambda w_d x_b \quad (1)$$

$$V_1(x_n) = \frac{\left(x_n - \frac{1}{2}\right)^2 AV_b}{4\left(\frac{1}{2} - x_{n2}\right)^2} \text{ and } V_2(x_n) = \frac{\left(x_n + x_{nl} - \frac{1}{2}\right)^2 (AV_b - E_g)}{4\left(\frac{1}{2} - x_{nl} - x'_{n2}\right)\left(\frac{1}{2} + x_{nl} - x'_{n2}\right)} \quad (2)$$

$$T(E_e, x_{nl}, x_{n2}) = \exp\{-F(T_{11} - T_{22})\} \quad (3)$$

Where,

$$F = \left(\frac{2m_0}{\hbar^2}\right)^{\frac{1}{2}}$$

$$T_{11} = \frac{AV_b}{(2 - 4x_{n2})}\left[\left(x_{nl} + \frac{x_{nl}}{\lambda} + \frac{1}{2}\right)\sqrt{\left(x_{nl} + \frac{x_{n2}}{\lambda} + \frac{1}{2}\right)^2 - \frac{4E\left(\frac{1}{2} - x_{nl}\right)^2}{AV_b}} - \left(\frac{x_{n2}}{\lambda} + \frac{1}{2}\right)\sqrt{\left(\frac{x_{n2}}{\lambda} + \frac{1}{2}\right)^2 - \frac{4E\left(\frac{1}{2} - x_{nl}\right)^2}{AV_b}}\right]$$

and

$$T_{22} = \frac{E\left(\frac{1}{2} + x_{n2}\right)}{\sqrt{AV_b}} \ln \frac{\left(\frac{1}{2} + \frac{x_{n2}}{\lambda}\right)\sqrt{AV_b} - \sqrt{\left(\frac{1}{2} + x_{n2}\right)^2 \frac{AV_b}{4} - 4E\left(\frac{1}{2} - x_{n2}\right)^2}}{\left(x_{nl} + \frac{x_{n2}}{\lambda} + \frac{1}{2}\right)\sqrt{AV_b} - \sqrt{\left(x_{nl} + \frac{x_{n2}}{\lambda} + \frac{1}{2}\right)^2 AV_b - 4E\left(\frac{1}{2} - x_{n2}\right)^2}}$$

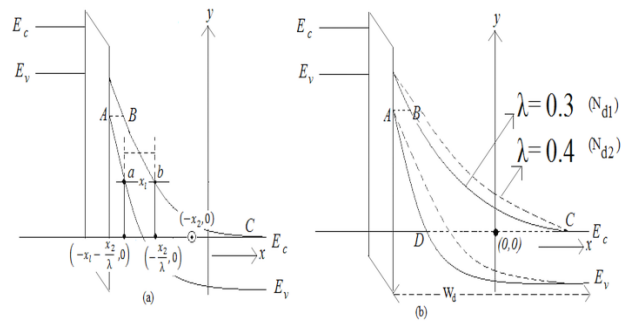

Fig.-2: Band diagram from BTBT Tunneling. (a) Nd2>Nd1 and it indicates that band banding decreases with increasing impurity doping and once it goes below to the Si band gap BTBT process will stop. (b) Compatibility of our model with band diagram AB and A'B' are pairs of classical tunneling points and its value is varied with incoming electron energy and impurity doping.

The validity of the proposed tunneling probability expression is guaranteed by the following plots (Fig-3a and Fig-3b) where T is varied with incoming Electron energy and tunneling width. Now from Fig-2b we have two boundary conditions:

$$at \; x_n = \lambda x_{n2} \quad ; \quad V_{n1}(x_n) = E_e \quad (4)$$
$$and \; x_n = \lambda x'_{n2} \quad ; \quad V_{n2}(x_n) = E_e \quad (5)$$

Now by solving equation (2), (4) and (5) we can get the expressions for $x_{n2}$ and $x'_{n2}$ in terms of $\lambda$ and $E_g$. Now the normalized tunnel length is $x_{nl} = x_{nl}(E, \lambda) = x_{n2} - x'_{n2}$. The dependency of this length on incoming electron energy and $\lambda$ is shown in Fig3. As $\lambda$ increases the variation of tunnel length decreases more rapidly (Fig:3c) with the increase of electron energy which is matched with diagram of Fig 2 and in other word it can be said that electrons with higher potential energy will have to tunnel a longer path (fig: 3d). From base of these results we can ensure that our tunneling model is totally valid. The resultant current density can be related to transmission probability as following,

$$I_{GIDL}\big|_{T_{GIDL}} = A_d J(E_e) = A_d \frac{q m_0}{2\pi^2 \hbar^3} T(E_e, V_b) \int_{E_e}^{\infty} \{f_v(E) - F_c(E)\} dE \quad (6)$$

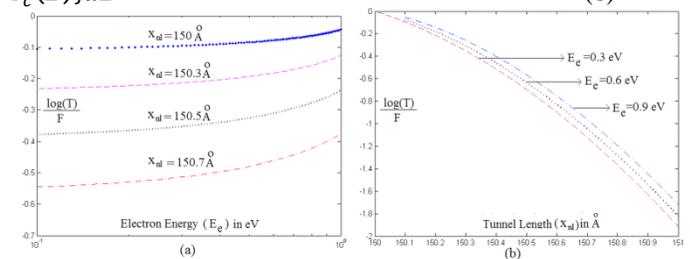

Fig.3: Variation of Tunneling Probability with (a) Electron Energy and (b) Tunneling Length according to our Model.

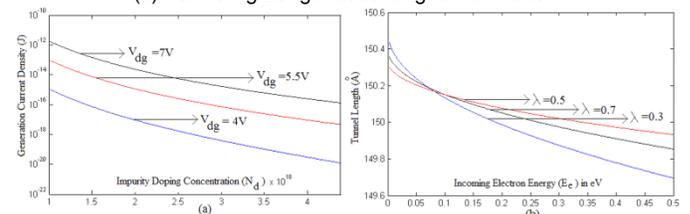

Fig.4: Effect of (a) Increasing and (b) Decreasing Impurity Doping Concentration

Where $m_0$ is the rest mass of electron, $\hbar$ is reduced plank constant and $A_d$ is the effective area. $f_v$ and $f_c$ are probability distribution function of valence and conduction band respectively. By assuming valence band full of electrons and conduction band with no electron we can put $f_v=1$ and $f_c=0$.
Figure 4(a) explained that electron-hole pair generation is decreases with $N_d$ increases and after band-bending falls below 2.1eV generation will stop and from Figure-4(b) it is clear that electron energy level and tunnel length are inversely proportional and the degree of dependency is controlled by $\lambda$ which is also inversely proportional with $N_d$.

## IV. Effect of Impurity Doping Concentration

To understand the effect of increasing of doping concentration we have to take a close look of fig: where the area ABCD is decreased with band bending which strongly dependent on impurity doping. By integrating the potential barrier (v2) curve with respect to x we can extract the value of area which has the physical interpretation of nothing but applied electric field.

$$\varepsilon_{GIDL} = \int_0^{W_d} V_2(x) = \int_0^{W_d} \frac{\left(x + x_l - \frac{1}{2}\right)^2 (AV_b - E_g)}{4\left(\frac{1}{2} - x_l - x_2'\right)\left(\frac{1}{2} + x_l - x_2'\right)} dx$$

$$or, \varepsilon_{GIDL} = 92.5 \cdot \left(W_d - \frac{1}{2}\right)^3 (V_b - E_g) \text{ and } W_d = \sqrt{\frac{2\epsilon_s V_b}{qN_d}}$$
(7)

$$I_{GIDL} = A\varepsilon_{GIDL}^2 e^{-\frac{B}{\varepsilon_{GIDL}}}$$

Now $I_{GIDL}|_{Total} = f(I_{GIDL}|_{\varepsilon_{GIDL}}, I_{GIDL}|_{T_{GIDL}})$
or, $I_{GIDL}|_{Total} = \alpha(I_{GIDL}|_{\varepsilon_{GIDL}}) \cdot \beta(I_{GIDL}|_{T_{GIDL}})$
or, $I_{GIDL}|_{Total} = C \cdot (I_{GIDL}|_{\varepsilon_{GIDL}})(I_{GIDL}|_{T_{GIDL}})$ (8)

Where $C = \alpha \cdot \beta$ and Equation (8) is the final required equation of our proposed model is $N_d$ dependent.

## V. Results and Discussions

Using equation () the leakage current is plotted with impurity doping concentration with different drain to gate voltage and the results are approximately matched with Endou's model. The parameter $\lambda$ is ratio between one of the control point of the potential barrier equation to one of the classical tunneling point (). If that particular control point is dragged towards left side on x axis potential barrier becomes sharper at the initial but meets with valence band top level at the same point.

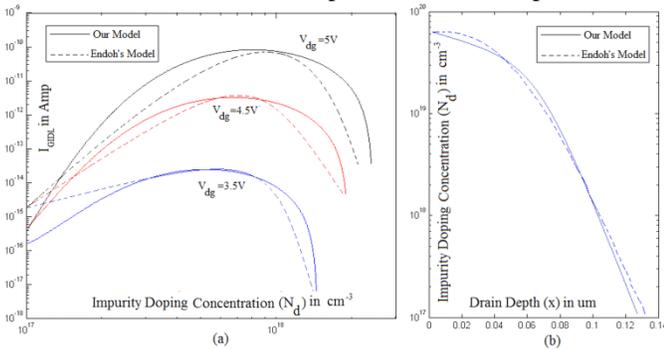

Fig.-5: (a) Variation of $I_{GIDL}$ with $N_D$ and (b) Drain Impurity Doping Profile where Drain Depth is actually the control point $x_2$ which is varying in x direction.

TABLE – I
DETAILS OF MODEL PARAMETERS WITH VALUES

| Model Parameters | Interpretation | Values |
|---|---|---|
| A | Proportionality Constant | 400 |
| C | Fitting Parameter of Our Model | $9.25 \times 10^{-11}$ |
| $\lambda$ | Proportionality Constant | 0 to 1 |
| $\gamma$ | Impurity Gradient | $10^{18}$ |

Thus this control point can be related to the impurity doping concentration with the following relation $N_d = \gamma x_2$. We have used famous Kane's model to get the expression of generation current $(I_{GIDL}|_{\epsilon_{GIDL}})$ in terms of impurity doping concentration. The corresponding plot is shown in figure. Interestingly this $\gamma$ have same unit as impurity gradient ($cm^{-4}$) as expected and corresponding plot of our software, given in fig-5(b) which is also approximately matched with the 2D process simulator used in Endoh's model. From Fig.-5(b) and Equation (1) it is also clear that $N_d$ and $\lambda$ are inversely proportional as assumed previously. In equation (7) if $V_b$ approaches to $E_g$ effect of electric field is reduced which results less $I_{GIDL}$ for low impurity doping concentration as mentioned in theory. Thus our model explains almost each and every background theory in geometrical way and also approximately matched with the previous model as shown in fig-5 and only one fitting parameter is used. In a nutshell the proposed model is well acceptable and has a right to claim novelty for the uniqueness.

## VI. Conclusion

The dependency of leakage current upon impurity doping concentration is explained both analytically and physically in details. Results of the proposed model and drain impurity doping profile are approximately matched with the previous model.

## VII. Appendix

Equation Formation from Bezier Curve Control Points:
$$P_0 \equiv [x_1, y_1], P_1 \equiv [x_2, y_2], P_3 = [x_3, y_3]$$
Now, the equations formed by the control points are as follows
$$x = (1-t)^2 x_1 + 2(1-t)t x_2 + t^2 x_3$$
And $y = t^2 y_3$
After eliminating t and neglected the terms where $y_3$ appears in denominator we get,
$$y \cong \frac{(x - x_1)^2 y_3}{4(x_2 - x_1)^2}$$

Here the Bezier Curves are used to model n band diagram. Thus obviously x and y axis denotes distance (in micro order) and energy level (in eV order) respectively. Thus either $y_3$ or $Ay_3$ appears in denominator and the numerator contains x or x axis related terms the factor can be easily neglected.

## VIII. References

[1] J. Chen, T. Y. Chan, I. C. Chen, P. K. KO, and Chenming Hu, "Subbreakdown Drain Leakage Current in MOSFET," *IEEE Electron Device Letters*, VOl. EDL-8, No. 11, November 1987, pp 515-517.

[2] T. Endoh, R. Shirota, M. Momodomi and F. Masuoka," An accurate model of subbreakdown due to band-to-band tunneling and some applications," *IEEE Trans. on Electron Devices*, Vol. 37, Jan 1990, pp. 290.

[3] Kuo-Feng You, Ching-Yuan Wu," N New Quasi-2-D Model for Hot-Carrier Band-to-Band Tunneling Current," *IEEE Transaction on Electron Devices*, Vol.46, No. 6, June 1999 pp. 1174-1179.

[4] Ja-Hao Chen, Shyh-Chyi Wong, and Yeong-Her Wang,"An Analytic Three-Terminal Band-to-Band Tunneling Model on GIDL in MOSFET," *IEEE Trans. on Electron Devices*, vol. 48, No. 7, July 2001, pp. 1400 – 1405.

[5] Xiaoshi Jin, Xi Liu and Jong-Ho Lee,"A Gate-Induced Drain-Leakage Current Model for Fully Depleted Double-Gate MOSFETs," *IEEE Trans, on Electron Devices* vol. 55, No 10, October 2008, pp. 2800-2804.

[6] Farkhanda Ana and Najeeb-ud-din,"SUPPRESSION OF GATE INDUCED DRAIN LEAKAGE CURRENT(GIDL) BY GATE WORKFUNCTION ENGINEERING: ANALYSIS AND MODEL," *Journal of Electron Devices*, vol. 13, 2012, pp. 984-996

[7] Ben G. Streetman and Sanjay Kumar Banerjee," Field-Effect Transistors," *SOLID STATE ELECTRONIC DEVICES*, 6th ed.Delhi-India, PHIL Ch. 6, Sec. 6.5.12, pp. 323-325.

[8] E. Yashida, T. Tanaka," A capacitor less 1T-DRAM technology using Gate-Induced Drain-Leakage (GIDL) current for low-power and high speed embedded memory," *IEEE. Trans. Of Electron Devices*, vol. 53 No.4 April -2006 pp. 692-697

[9] Te-Chih Chen, Ting-Chang Chang, Fu-Yen Jian, Shih-Ching Chen, Chia-Sheng Lin, Ming-Hsien Lee, Jim-Shone Chen, Ching-Chieh Shih," Improvement Memory State Misidentification Caused by Trap-Assisted GIDL Current in a SONOS-TFT Memory Device," *IEEE Trans. On Electron Device Letters*.Vol. 30 no. 8, August – 2009 pp. 834-836.

[10] M. Sarfraz, M. Riyazuddin, M. H. Baig, " Capturing Planner shapes by approximating their outlines," *Journal of Computational and Applied Mathematics*, vol. 189 issue 1-2 May 2006, pp. 494-512.

[11] Y. J. Ahn," Conic Approximation of Planner Curves," *Computer-Aided Design*, vol. 33, issue 12, October 2001, pp. 867-872.

[12] E. B. Abbott, M. Lee, R. S. Mand, M. Sweeny, J. M. Xu," A quantum gate current model." *IEEE Trans. On Electron Devices* vol.40 no-5 May 1993 pp.1022-1024.

[13] Clarence R. Crowell and MadjidHafizi,"Current Transport over Parabolic Potential Barriers in Semiconductor Devices," *IEEE Trans. on Electron Devices*, vol. 35 No. 7, July-1988, pp. 1087-1095.

[14] A. Sen, J. Das ," MOSFET GIDL Currentby Designing a New BTBT Model Using De-Casteljau's Algorithm " unpublished.

[15] Sanjay Kumar, EktaGoel, Kunal Singh, Balraj Singh, Mirgender Kumar and SatyabrataJit, "N Compact 2-D Analytical Model for Electrical Characteristics of Double-Gate Tunnel Field-Effect Transistors With n $SiO_2$/High-k Stacked Gate-Oxide Structure," *IEEE Trans. on Electron Devices*, vol. 63, no. 8, August – 2016, pp. 3291-3299.

[16] Yuji Ando and Tomohiro Itoh,"Calculation of transmission tunneling current across arbitrary potential barriers," *Journal of Applied Physics*, vol. 61, no. 4, February – 1987, pp. 1497-1502.

[17] Pradipta Dutta, Kalyan Loley, Arka Dutta, Chandan Kumar Sarkar, " An Analytical BTBT Current Model of Symmetric/ Asymmetric 4T Tunnel Double Gate FETS With Ambipolar Characteristics," *IEEE Trans. On Electron Devices*, vol. 63 no. 7 July 2016 pp. 2700-2707

[18] Yunhe Guan, Zunchao Li, Wenhao Zhang, Yefei Zhang, Feng Liang, "An Analytical Model of Gate-All-Around Heterojunction Tunneling FET,"*IEEE Trans. On Electron Devices*, vol. 65 no.2, Feb 2018 pp. 776-782.

[19] E. O. Kane, "Zener tunneling in semiconductor," *J. Phys. Chem. Solids*, vol 12, 1959 p. 181

[20] S. M. Sze, *Physics of Semiconductor Devices*, 2nd ed. New York: Wiley, 1981, p. 81